\documentstyle[twocolumn,aps,prd,eqsecnum,epsfig]{revtex}
\tighten

\def\bfl{\begin{flushleft}}
\def\efl{\end{flushleft}}
\def\bfr{\begin{flushright}}
\def\efr{\end{flushright}}
\def\bc{\begin{center}}
\def\ec{\end{center}}
\def\be{\begin{equation}}
\def\ee{\end{equation}}
\def\ba{\begin{eqnarray}}
\def\ea{\end{eqnarray}}
\def\nn{\nonumber }
\def\lb#1{\label{#1}}

\def\drm{d}

\def\schrod{Schr\"odinger  }

\def\rphi{X}
\def\phik#1#2{\rphi_{#1}^{(s)\,#2}}
\def\Mass{{\cal M}}
\def\gaug{\varphi}
\def\by{\ast}

\def\Det#1{\, \text{det}\left(#1 \right) }
\def\Sign#1{\, \text{sign}\left(#1\right) }
\def\PDer#1#2{\,\frac{\partial #1}{\partial #2}}

\def\Sum#1#2{\, \sum\limits_{#1}^{#2}}
\def\ftr{\, \verb|!|}

\def\Cosh#1#2{\, \text{cosh}^{#1}\left(#2 \right) }
\def\Sinh#1#2{\, \text{sinh}^{#1}\left(#2 \right) }
\def\Tanh#1#2{\, \text{tanh}^{#1}\left(#2 \right) }
\def\Ln#1#2{\, \text{ln}^{#1} \left(#2 \right) }
\def\Expi#1{\, \text{e}^{#1} }

\begin{document}

\wideabs{
\draft

\title{
\footnotesize
~~~~~~~~~~~~~~~~~~~~~~~~~~~~~~~~~~~~~~
~~~~~~~~~~~~~~~~~~~~~~~~~~~~~~~~~~~~~~
~~~~~~~~~~~~~~~~~~~~~~~~~~~~~~~~~
hep-th/0008004\\
~~~~~~~~~~~~~~~~~~~~~~~~~~~~~~~~~~~~~~
~~~~~~~~~~~~~~~~~~~~~~~~~~~~~~~~~~~~~~
~~~~~~~~~~~~~~~~~~~~~~~~~~~~~~~~
Phys. Lett. B 519 (2001) 111-120  [Secs II-IV only]\\
~\\
\large \bf
Holography and field-to-particle transition 
in sigma model,
dilaton and
supergravity:
Point particle
as end product of field theory
and seed of string-brane approach
      }

\author{Konstantin G. Zloshchastiev}

\address{
Department of Physics, National University of Singapore,
Singapore 119260, Republic of Singapore\\
and 
Department of Theoretical Physics, Dnepropetrovsk State University,
Dnepropetrovsk 49050, Ukraine
}

\date{~Received: 1 Aug 2000 [LANL], 19 Jul 2001 [PLB] ~}
\maketitle

\begin{abstract}
The paper consists of the two parts that are devoted to 
separate problems and hence seem to be independent of
each other but for a first look only.
In the Part one the field-to-particle transition formalism
is applied to the sigma model (field counterpart of string theory), 
dilaton gravity and gauged supergravity 0-brane solutions.
In addition to the fact that the field-to-particle transition is of interest
itself it can be regarded also
as the nontrivial dynamical dimensional reduction which takes into account
field fluctuations as well as one can treat it as
the method of the consistent
quantization in the vicinity of the nontrivial vacuum
induced by a field solution.
It is explicitly shown that in all the cases the end product of
the approach is the so-called non-minimal 
point particle.
In view of this 
it is conjectured that this is not a mere coincidence, and
point particles are not only the
end product of the field theory but also the underlying base of 
material world's hierarchy:
if high-dimensional point particles are the base
then the
strings can be regarded as their path trajectories, etc.
The Part two is devoted to the formal axiomatics 
of the string-brane approach -
there we deepen the above-mentioned conjecture and make it more accurate.
It is shown
that a point particle 
can be regarded as an extended object from
the viewpoint of a macroscopic measurement -
it is actually observed as the
``cloud'' consisting of the particle's virtual paths,
therefore, for an observer the real particle cannot be  
seen  as a point object anymore.
However, it cannot be supposed also as a continuous extended
object unless we average over all the deviations.
Once we have performed the averaging we obtain the 
extended object,  
metabrane, which can be considered ``as is'', i.e.,
regardless of the underlying microstructure because 
it possesses the new set of the (collective) degrees of freedom.
This brane may appear to be the composite string-brane object
consisting of the microscopical objects (strings) and cosmological-size
ones
(3-brane) where the  size of a corresponding embedding
is governed by the  weight constants 
arising along with the decomposition.
Finally, we outline the picture which unifies both parts and
sets the role of each of them.
\end{abstract}

\pacs{PACS number(s): 04.60.Kz, 04.65.+e, 11.25.Sq, 11.27.+d, 12.60.Jv}

}


\narrowtext

\section{Introduction}\lb{s-i}

In the part one of 
present paper we try to achieve the three main aims:
(i) 
construction of the consistent classical and 
quantum field-to-particle transition formalism for the
$N$-component $2D$ sigma model,
(ii) 
application of this approach to classical and 
quantum dilaton gravity,
(iii) application  to 
supergravity 0-brane solutions and connection with the
holographic principle.
Let us consider them in more details.

{\it Field-to-particle transition}.
The study of the field-to-particle transition formalism itself
lies within the well-known dream program of constructing a 
theory which would not contain matter as external postulated entity but
would consider fields as sources of matter and particles as
special field configurations.
Such a program was inspired probably 
first by Lorentz and Poincar\'e and since that time many efforts have been
made to come it into reality, one may recall the Einstein's, Klein's
and Heisenberg's attempts.
The discovery of fermionic fields and success of Standard Model
decreased interest to such theories but,
nevertheless, the problem of the origin of matter 
remains to be still open and important, 
especially in what concerns the theoretical explanation of
the fundamental properties of observable particles.

The most intuitive idea here is to regard the localized 
field solutions as the corpuscles of matter. 
However, if one tries to fit such a field solution in the particle 
interpretation one eventually encounters the problem of how to treat
the field fluctuations in terms of particle degrees of freedom.
In fact, it is caused by the fact that a
field solution is still a field and hence it
has an infinite number of  degrees of freedom
whereas a particle has a finite number of them.
Therefore, it is necessary to correctly handle this circumstance
to be protected from deep contradictions.
Thus,
in Sec. \ref{s-ea-gf} we demonstrate the
field-to-particle transition for the
$N$-component $2D$ sigma model or, equivalently, simple bosonic
string in the $N$-dimensional spacetime (the latter is supposed flat
for simplicity but the external potential $U$ is included and can
be treated as effective).
Minimizing the action with respect to field fluctuations around
a chosen solution,
we remove zero modes and eventually obtain the 
effective point-particle action where
the non-minimal curvature-dependent ({\it rigidity}) 
terms \cite{txt-nm} are caused by the 
fluctuations in the vicinity of the solution.  
Sec. \ref{s-ea-qu} is devoted to quantization of the 
effective action as a constrained
mechanics with higher derivatives.
There we discuss the recent state of the theory and
we calculate the quantum corrections to the
mass of such a field/string-induced particle.

{\it Field-to-particle transition and dilaton gravity}.
It is well-known that the black holes and singularities
arise in gravity as field solutions of the corresponding equations
of motion.
On the other hand, from the viewpoint of general
relativity the point particle can be regarded as the pointwise
singularity -
either non-coordinate and naked or hidden under horizon (black hole).
As a rule, people make no difference between the concepts 
``black hole as a field solution'' and
``black hole as a particle''.
However, in view of the previous paragraph the field solution and
naively associated particle are not the identical entities, 
and it may result in contradictions 
when considering the perturbational, statistical and quantum aspects of
the black hole theory.
The two-dimensional theory of gravity \cite{bnc} seems to be
a good place of demonstration of how the field nature of
a black hole should be properly taken into account.
The success of the field-to-particle approach in the $2D$
dilaton gravity is very important, especially, if one recalls
that the latter
can be viewed also as a dimensional reduction of the 3D BTZ black
hole \cite{btz,ao} and spherically symmetric solution of 4D
dilaton Einstein-Maxwell gravity used as a model of
evaporation process of a near-extremal black hole \cite{cm}.

Using the property that the gauged dilaton gravity can be
regarded as a sigma model and keeping in mind the approach
developed in Sec. \ref{s-ea}, we demonstrate how for any
given localized 
dilaton-metric solution
(having the global properties of either a black hole or a naked
singularity) one can consistently perform the
field-to-particle transition (taking into account the fluctuations
of the dilaton field and metric) and obtain
the action of the nonminimal particle with the rigidity caused by
field fluctuations.
In fact, only then it is possible to
consistently quantize the field solution,
and we demonstrate how to do it in a standard and uniform way
within frameworks
of the (nonminimal) quantum mechanics of Sec. \ref{s-ea-qu},
emphasizing on the corrections to mass spectrum.
Besides, it should be noted that the main results are obtained in a 
non-perturbative way that seems to be important for highly non-linear 
general relativity. 

One may ask the following question:
there already exists the wide literature devoted to classical and quantum
aspects of the theory \cite{lgk,dil-qua}
so what are the advantages of the quantization based on 
the nonminimal-particle approach?
The answer is evident if one understands that the quantizations
in the vicinity of the trivial vacuum (the classical ground state is
zero or at least a constant) and in the vicinity of the
nontrivial vacuum (the ground state - classical field solution -
is some function of coordinates) differ from each other.
The latter is more complicated because of, first of all, the appearance of 
zero modes \cite{raj}.
Therefore, to correctly quantize a black hole it is necessary
to quantize the gravity in the  vicinity of such a nontrivial vacuum
($=$ classical black hole solution)
and handle the above-mentioned features.
The theory based on a field-to-particle transition,
where there is no rigid fixation of spatial symmetry and where
field fluctuations and zero modes are properly taken into account,
does it.

{\it Field-to-particle transition and holography}.
In Sec. \ref{s-sg} we demonstrate how 0-brane solutions of 
the general supergravity in arbitrary spacetime dimensions $D$
can be described
in terms of the mechanics of
non-minimal particles with rigidity as well as
how this circumstance appears to be in agreement with the holographic 
principle \cite{holo}
and, finally, how it is related to the AdS/CFT conjecture \cite{mal}.
The holographic principle in its most radical form suggests
that a bulk $d$-dimensional theory is equivalent to a boundary 
$(d-1)$-dimensional theory.                   
In our case the bulk theory is the two-dimensional
(gauged) supergravity whereas the corresponding boundary theory
will be shown to be given by the nonminimal mechanics
with rigidity.
The latter
is invariant under the transformations of the proper parameter $s$,
and
one may recall another diff-invariant 
candidates for the role of a
0-brane boundary theory,
the generalized conformal mechanics of a probe 0-brane 
\cite{cdkktv,you} and the De Alfaro-Fubini-Furlan model
coupled to an external non-integrable source \cite{cckm},
which fulfil the Maldacena's conjecture.
However, the probe mechanics by construction takes into account neither
the field fluctuations
near the brane solution nor arising zero modes whereas the roles of both
are very important, following the aforesaid and Sec. \ref{s-ea}.
If one properly considers these aspects when  performing
the consistent field-to-particle transition then
the appearance of non-minimal, depending on world-line
curvature, terms is inevitable.

In Sec. \ref{s-d} we discuss the part one (Secs. 
\ref{s-ea} - \ref{s-sg}) and make the conjecture
upon whether a point-particle theory could be the 
the underlying theory of the strings and branes which
are regarded now as the basis of modern high-energy physics.

Further, the successive Sec. \ref{s-ps} 
begins the (self-sufficient) part two
which aims to justify the proposed conjecture by putting  a
concrete physical mechanism forward.
There it is demonstrated how the taking of quantum uncertainty
into account leads to the ``transmutation'' of a point particle 
into the extended (brane-like) object.
In particular, we demonstrate the differences between the proposed
approach and the matrix theory \cite{bfss} which also exploits the
quantum point-particle concept but in quite different context.
Also we discuss the cosmological consequences 
in the context of the string-brane world.
                           
Final conclusions and unresolved problems 
are given in Sec. \ref{s-c} where we attempt
to unify the part one and part two into an integral picture
and clearly point out the role and the place of each of them.

\section{Effective nonminimal-particle action from sigma model
(string)}\lb{s-ea}

In this section we will construct the nonlinear effective action of 
the sigma model (bosonic string) in the vicinity of 
a localized solution, and then consider its quantum aspects.
In fact, here we will
describe the procedure of the correct transition from field 
to particle degrees of freedom.
Indeed, despite the localized solution resembles a particle
both at classical and quantum levels,
it yet remains to be a {\it field} solution with infinite number of
field degrees of freedom
whereas a true particle has a finite number of degrees of freedom.
Therefore, we are obliged to correctly handle this circumstance
(as well as a number of others)
otherwise deep contradictions may appear.

\subsection{General formalism}\lb{s-ea-gf}

Let us begin with the action of the following $N$-component 2D sigma model
\begin{equation}
S[\rphi] = \int  L(\rphi)\, \drm^2 x,
                                                            \lb{eq2.1} 
\end{equation}
where the Lagrangian is given by
\be                                                          \lb{eq2.2}
L (\rphi) = \frac{1}{2} 
\sigma_{a b} \partial_\nu \rphi^a \partial^\nu \rphi^b -
U (\rphi),
\ee
where $\sigma$ is a constant $N\times N$ matrix, 
Greek indices run from $0$ to $1$,
and $a,\, b = 1,\, ..., N$.
The corresponding equations of motion are
\be                                                          \lb{eq2.3}
\sigma_{a b}\, \partial^\nu \partial_\nu \rphi^b + U_a(\rphi) = 0,
\ee
where we have defined                                   
\[
U_a(\rphi) = \PDer{U(\rphi)}{\rphi^a},~~
U_{ab}(\rphi) = \PDer{^2\, U(\rphi)}{\rphi^a \partial \rphi^b}.
\]
Suppose, we have the solution which depends on the single 
combination of initial variables, e.g., the Lorentz-invariant one:
\be                                                           \lb{eq2.4}
\phik{a}{}(\rho) = \phik{a}{}
\left( \gamma ( x-v t)
\right),
~~\gamma= 1/\sqrt{1-v^2},
\ee
possessing localized Lagrangian density
in the sense that the mass integral over the domain of 
applicability is finite
\be                                                            \lb{eq2.5}
\mu = - \int\limits_{\rho_1}^{\rho_2}
L (\phik{}{})\ \drm \rho < \infty,
\ee
coinciding with the total energy up to the sign and 
Lorentz factor $\gamma$ 
(below the boundaries $\rho_i$
will be omitted for brevity).

Let us change to the set of the collective coordinates 
$\{\sigma_0=s,\ \sigma_1=\rho\}$ such that
\be  
x^m = x^m(s) + e^m_{(1)}(s) \rho,\ \              
\rphi_a(x,t) = \widetilde \rphi_a (\sigma),     
\ee
where $x^m(s)$ turn out to be the coordinates of a (1+1)-dimensional point
particle, $e^m_{(1)}(s)$ is the unit spacelike vector orthogonal
to the world line.
Hence, the initial action can be rewritten in new coordinates as
\be                                                         \lb{eq2.7}
S[\widetilde \rphi] = 
\int L (\widetilde \rphi) \,\Delta \ \drm^2 \sigma,
\ee
and
\[
L (\widetilde \rphi) = \frac{1}{2} 
\sigma_{a b}\, \partial_- \widetilde\rphi^a 
\partial_+ \widetilde\rphi^b
- U (\widetilde \rphi),
\]
where
$ \partial_\pm = \frac{1}{\Delta} \partial_s \pm \partial_\rho 
$, and
\[
\Delta = \text{det} 
\left|
\left|
      \PDer{x^m}{\sigma^k}
\right|
\right|
= \sqrt{\dot x^2} (1- \rho k), 
\]
whereas $k$ is the curvature of a particle world line
\be                                                            
k = \frac{\varepsilon_{mn} \dot x^m \ddot x^n}{(\sqrt{\dot x^2})^3},
\ee
where $\varepsilon_{m n}$ is the unit antisymmetric tensor.
This new action contains the $N$ redundant degrees of freedom which 
eventually
lead to appearance of the so-called ``zero modes''.
To eliminate them we must constrain the model
by means of the condition of vanishing of the functional derivative with
respect to field fluctuations about a chosen static solution,
and in result we will obtain the required effective action.

So, the fluctuations of the fields $\widetilde\rphi_a (\sigma)$ in the 
neighborhood of the static solution $\phik{a}{} (\rho)$
are given by the expression
\be                                               
\widetilde\rphi_a (\sigma) = 
\phik{a}{} (\rho) + \delta \rphi_a (\sigma).
\ee
Substituting them into the expression (\ref{eq2.7}) and 
considering the static
equations of motion for $\phik{a}{}(\rho)$ we have
(below we will omit the superscript ``$(s)$'' at $\rphi$'s for 
brevity assuming that all the values are taken on the solutions
$\phik{a}{} (\rho)$):
\ba                                                         \lb{eq2.10}
&&S[\delta \rphi] 
= \int d^2 \sigma \ 
   \Biggl[
        \frac{\Delta}{2} 
        \biggl(2 L +
                       \sigma_{a b} \,
                       \partial_-  \delta \rphi^a \,
                       \partial_+ \delta \rphi^b
\nn\\ &&           
                   - 
               U_{ab} \delta \rphi^a\, \delta \rphi^b
        \biggr) 
        +
        \Delta^\prime \,
        \sigma_{a b} \,  \rphi^{a \prime}  \,\delta \rphi^b
        + O (\delta \rphi^3)                            
    \Biggr] + \breve S,                                            
\ea
where  $\breve S$ are surface terms (see below for details),
\[
L = -
 \frac{1}{2} \sigma_{a b} \, \rphi^{a \prime} \rphi^{b \prime} - U,
\]
and prime means the derivative with respect to $\rho$.
Extremizing this action with respect to 
$\delta \rphi_a$ one can obtain 
the system of equations in partial derivatives for field fluctuations:
\ba
&&
\left(
     \partial_s \Delta^{-1} \partial_s -
     \partial_{\rho} \Delta \partial_{\rho} 
\right) \delta\rphi^a
+\Delta 
\sigma^{a c}
U_{c b} \delta\rphi^b \nn\\
&& \qquad\qquad\qquad\qquad\quad + \rphi^{a \prime} \, k\sqrt{\dot{x}^2} +
O(\delta \rphi^2) = 0,
\ea
which is the constraint removing redundant degrees of
freedom, besides we have denoted the inverse matrix as $\sigma$
with raised indices.
Supposing $\delta\rphi_a (s,\rho) = k(s) f_a(\rho)$, in the 
linear approximations
$\rho k\ll 1$ (which naturally guarantees also
the smoothness of a world line at $\rho \to 0$, i.e., the absence of 
a cusp in this point) 
and $O(\delta\rphi^2)=0$ we obtain the system
of three ordinary derivative equations
\ba  
&&\frac{1}{\sqrt{\dot{x}^2}} \frac{d}{ds} 
\frac{1}{\sqrt{\dot{x}^2}} \frac{dk}{ds} +ck = 0,          \lb{eq2.12}\\
&&
-f^{a \prime \prime} + 
\left(
      \sigma^{a c} U_{c b} - c \delta^a_{b}
\right) 
f^b + \rphi^{a \prime} = 0,                       
\ea
where $c$ is the constant of separation.
Searching for a solution of the last subsystem in the form 
(we assume that we can have several $c$'s and thus no summation over 
$a$ in this and next formula)
\be                                                       \lb{eq2.14}
f_a = g_a + \frac{1}{c_{\hat a}} \rphi^{\prime}_a,
\ee
we obtain the homogeneous system
\be  
-g^{a \prime\prime } + 
\left(
      \sigma^{a c} U_{c b} - c_{\hat a} \delta^a_{b}
\right)
g^b = 0.                                           \lb{eq2.15}
\ee
Strictly speaking, the explicit form of $g^a (\rho)$ is not significant 
for us, because we always can suppose integration constants to be zero
thus restricting ourselves by the special solution.
Nevertheless, the homogeneous system should be considered as the 
eigenvalue problem for $c_{\hat a}$'s (see below).
After resolving this eigenvalue problem  and 
substituting the found functions $\delta\rphi_a = k f_a$ and 
eigenvalues $c_{\hat a}$
back in the action (\ref{eq2.10}), 
we can rewrite it in the explicit zero-brane form (omitting surface
terms)
\be                                
S_{\text{eff}} = 
S_{\text{eff}}^{\text{(class)}} + S_{\text{eff}}^{\text{(fluct)}} =
- \int \drm s \sqrt{\dot x^2} 
\left(
       \mu + \alpha_2^{(0)} k^2
\right),                                 \lb{eq2.16}
\ee
describing the nonminimal point-particle with curvature,
where $\mu$ is the value of the mass integral above, and
\be                                               
\alpha_2^{(0)} = 
\frac{1}{2} \int\limits^{\rho_2}_{\rho_1} \sigma_{a b} \,  
 \rphi^{a \prime} f^{b} \, \drm \rho = \frac{\mu}{2 c_{\hat a}}.
\ee
Let us recall the surface terms $\breve S$ in the action (\ref{eq2.16}).
In general case they are non-zero and give a 
contribution:
\[                                
S_{\text{eff}} = 
- \int \drm s \sqrt{\dot x^2} 
\left(
       \mu + \alpha_1 k + \alpha_2 k^2 + \alpha_3 k^3
\right),
\]
where 
\ba
&& \alpha_1 = \check\alpha_1 \bigl|^{\rho_2}_{\rho_1} ,\ \
\check\alpha_1 \equiv \sigma_{a b} \rphi^{a \prime} f^b , \nn\\
&& \alpha_3 = 
   - \frac{1}{2} \rho\, \sigma_{a b} f^{a \prime} f^b
     \biggr|^{\rho_2}_{\rho_1}, \nn\\
&& \alpha_2 = 
\alpha_2^{(0)} + \alpha_3 - 
(\rho \check\alpha_1)\biggr|^{\rho_2}_{\rho_1}.  \nn
\ea
However, 
the natural requirement of vanishing of field fluctuations
at the spatial boundaries leads to the canceling of
the terms containing $k$ and $k^3$.
Thus, below we will regard the action (\ref{eq2.16}) with 
$\breve S = 0$ as a basic one.
It is straightforward to derive the corresponding 
equation of motion in the Frenet basis
\be
\frac{1}{\sqrt{\dot x^2}}
\frac{\drm}{\drm s}
\frac{1}{\sqrt{\dot x^2}}
\frac{\drm k}{\drm s} +
\left(c - \frac{1}{2} k^2
\right) k = 0,
\ee
hence one can see that eq. (\ref{eq2.12}) was nothing
but this equation in the linear approximation with respect
to curvature,
as was expected.

Thus, the only problem, which yet demands on 
resolving is the determination of the concrete (eigen)value of
the constant(s) $c$.
It is equivalent to the eigenvalue problem for the system 
(\ref{eq2.15}) under some chosen boundary conditions.
If one supposes, for instance, the finiteness of $g$ at infinity
then the $c$'s spectrum turns out to be discrete.
Moreover, it often happens that $c$ has only one or two admissible
values if the effective potential is steep (e.g., of exponential type).
In any case, the exact value of $c$ is necessary hence the system 
(\ref{eq2.15}) should be resolved as exactly as possible.
Let us consider it more closely.
The main problem is the functions $g_a$ are mixed between
equations.
To separate them, let us recall that there exist $N-1$ orbit
equations, whose varying resolves the separation problem.
Let us demonstrate this for the most important for us 
case $N=2$.
Considering Eq. (\ref{eq2.14}), the varying of a single 
orbit equation yields
\be
\frac{\delta \rphi_2}{\delta\rphi_1} = 
\frac{\rphi_{2}^{\prime}}{\rphi_{1}^{\prime}} =
\frac{g_2}{g_1},
\ee
hence the system (\ref{eq2.15}) at $N=2$ can be separated into the
two independent equations
\be  
-g''_a + 
\left( 
      \frac{\rphi_{a}^{\prime\prime\prime}}
           {\rphi_{a}^{\prime}}
      - c_{\hat a}
\right) g_a = 0,                                           \lb{eq2.20}
\ee
if one uses 
$\displaystyle{\rphi^{a\prime\prime\prime} = 
\sigma^{a c} U_{c b} \rphi^{b \prime}}$.
In this form it is much easier to resolve the eigenvalue problem.
Therefore, the two independent parameters for the action (\ref{eq2.16}), 
$\mu$ and $c_a$, can be determined immediately by virtue of
eqs. (\ref{eq2.5}) and (\ref{eq2.20}).

Thus, in this section we have constructed the 
effective  action for the sigma model in
the vicinity of a static solution taking into account
the field fluctuational corrections which eventually lead to the appearance
of nonminimal terms in the action.

Before we proceed to the next section let us answer the 
following related question which can arise:
well, now we have obtained some effective action which interprets
the localized soliton/solitary-wave field solution as a particle 
taking into account field fluctuations...
Can we learn something useful from this interpretation in addition
to the knowledge of this analogy itself?
To answer it let us recall that the quantizations
in the vicinity of the trivial vacuum (the classical ground state is
zero or at least a constant) and in the vicinity of the
nontrivial vacuum (the ground state - classical field solution -
is some function of coordinates) differ from each other:
the latter is more complicated, first of all, because of the appearance of 
zero modes \cite{raj}.
Therefore, to correctly quantize a field theory 
(sigma model, in our case)
in the  vicinity of such a nontrivial vacuum it is necessary
to handle the above-mentioned features.
The theory based on a field-to-particle transition
does handle them all and in result yields the effective particle action
which is eligible for further study including quantization.

\subsection{Hamiltonian structure and Dirac quantization}\lb{s-ea-qu}

The constructing of Hamiltonian and quantum pictures of
the nonminimal theories is a separate large problem.
From the particle-with-rigidity action 
\be                                               \lb{eqNMPAgen}
S = 
- \mu\int \drm s \sqrt{\dot x^2} 
\left(
       1 + \frac{1}{2 c} k^2 
\right),
\ee
and definition of the world-line curvature 
one can see that we have the 
theory with higher derivatives.
Besides, the Hessian matrix constructed 
from the derivatives with respect to accelerations,
\[
M_{a b} = 
\left|
\left|
\PDer{^2 L_{\text{eff}}}{\ddot x^a \partial \ddot x^b}
\right|
\right|,
\] 
appears to be singular that says about the presence of the
constraints on phase variables of the theory.
The constructing of Hamiltonian formalism 
and quantization of the quadratic-curvature
model has been performed, reperformed or improved by many
people \cite{ply88,pav,ply,kpp,zlo015}, see Chapter IVA of
Ref. \cite{zlo015} for a recent state of the theory and details.
Summarizing those results one can say that, despite the Hamiltonian
picture of phase variables and constraints is more or less understood,
some questions, 
such as the reliable (unique) 
obtaining of the mass spectrum
of such a non-minimal particle, remain to be open.

At least, the two following problems seem to be presented here.
At first, there appears no uniquely defined \schrod equation and 
hence it is known no unique spectrum of eigenvalues. 
This is caused by the gauge freedom which arises at the constructing
of the Hamiltonian by virtue of the Dirac approach.
When considering constraints on the phase variables we eventually
have the following set of constraints (the
definitions of Ref. \cite{ply} are used)
\ba
&& P q + \sqrt{- q^2} 
\left\{
      \mu + \alpha [ {\bf p}^2 + ({\bf p r})^2 ] 
\right\} \approx 0,            \nn    \\
&& \frac{P q}{\sqrt{-P^2}} + f ({\bf q}^2)  \approx 0,   \nn \\
&& \sqrt{-P^2} - \Mass \approx 0, \nn
\ea
where $\alpha = - c/\mu$ and $f > 0$ is some function which
depends on a concrete gauge: e.g., $f=\sqrt{1+{\bf q}^2}$ corresponds
to the proper-time gauge.
This function does not affect the Hamiltonian picture but 
appears in the final \schrod equation ($\hbar =1$), 
\[
- \Psi''(r) +
\left[
\frac{l (l+1)}{r^2 (1+r^2)}
+
\frac{\mu - \Mass\tilde f (r^2)}{\alpha (1+r^2)}
\right] \Psi (r) = 0,
\]
where ${\bf r}={\bf q}/\sqrt{-q^2}$,
$\tilde f = f / \sqrt{-q^2}$, $\Mass$ is the total mass,
$l$ is the spin
eigenvalue, a non-negative integer 
(half-integer? see the corresponding discussion and
criticism by Plyushchay \cite{ply} of 
the Pav\v si\v c's paper from Ref. \cite{pav}).
At second, there is no clear understanding what is the difference
between theories with positive and negative parameters $\alpha$ and $\mu$,
in particular, it is unclear at which values of those
parameters we have the discrete spectrum of $\Mass$ and
at which ones we have the continuous spectrum.
In fact, it is also caused by the first problem, i.e., by the
non-uniqueness of the \schrod equation in coordinate representation.
It should be pointed out that there exists the rather different
form of the \schrod equation \cite{zlo015}
whose appearance was caused by slightly different (from Plyushchay's)
way of constructing
of the Hamiltonian with constraints.
In this paper we will follow the latter approach because
its eigenvalue structure is well-studied and confirmed both by numerical
methods and by comparison with others approaches.
Besides, we are mainly interested in the quantum
corrections to the canonical mass of a solution so in this
connection we do the following.
Referring a reader for details to the Chapter IVA of
Ref. \cite{zlo015}, we begin here with some ready results.

Namely, it was shown that the first-order 
(with respect to Planck constant) corrections to the bare mass $\mu$
are given by the equation
\be                                                   \lb{eqMcal1}
2 \varepsilon = \sqrt{B (B-1)} + O (\hbar^2),
\ee
where 
\[
\varepsilon = 8 \mu^2 \Delta/c, \ \Delta=1-\Mass/\mu, \
B = 8 \sqrt{\mu \Mass / c},
\]
where all the values should be real ($B$ should be positive)
to ensure the existence of 
bound states and hence to guarantee
the  stability of the quantum nonminimal particle.
Further, Eq. (\ref{eqMcal1}) can be rewritten as 
\be                                                      \lb{eqMcal2}
\Delta - \frac{\sqrt c}{\mu}
\frac{ \sqrt[4]{1-\Delta}  }{ 4 \sqrt 2}
\sqrt{
       8 \sqrt{1-\Delta} -  \frac{\sqrt c}{\mu}
     }
 = O (\hbar^2).
\ee
Suppose first that $c$ and $\mu$ are positive.
Then the  ground-state
eigenvalue of $\Mass$ is given by $\Delta = 0$,
i.e.,  $\Mass_0 = \mu$.
We have deal with small oscillations around a ground state hence
expanding the Eq. (\ref{eqMcal2}) in the Taylor series in the vicinity
of the ground state $\Delta_0 = 0$ we eventually obtain
the desired correction
\be                                                        \lb{eqMcal3}
\frac{\Mass}{\mu} = 1 - 4 
\frac{a^2 (a^2 -16)}{a^4-64 a - 256} + O (\hbar^2 \Delta^2),
\ \ c>0,
\ee
where $a \equiv \sqrt 2 \sqrt{8 - \sqrt c/\mu}$.
Note, here unlike Ref. \cite{zlo015} we are not needed in further
approximations like $O (c/\mu^2)$ etc.

Now, what is happening when $c<0$ ?
The first impression is that imaginary terms appear 
in Eq. (\ref{eqMcal2}) and break bound states.
However, numerical simulations show that it is not so.
To demonstrate it analytically let us look at Eq. (\ref{eqMcal1})
and recall 
that our theory is defined up to the sign of $\mu$ (at least),
therefore, the substitution $\mu = - |\mu|$ and $c= - |c|$ 
(keeping $\Mass$ positive)
changes nothing in Eq. (\ref{eqMcal1})  (note,
the root in the R.H.S of this
equation is 
defined up to a sign), and
eventually we have 
instead of Eq. (\ref{eqMcal2}) the following one
\be                                                      \lb{eqMcal2n}
\bar\Delta - \frac{\sqrt{|c|}}{|\mu|}
\frac{ \sqrt[4]{\bar\Delta-1}  }{ 4 \sqrt 2}
\sqrt{
       8 \sqrt{\bar\Delta-1} -  \frac{\sqrt{|c|}}{|\mu|}
     }
 = O (\hbar^2),
\ee
where $\bar\Delta = 1 + \Mass/|\mu| $.
Note, now the ground state belongs to the lower 
continuum $\Mass_0 = -|\mu|$ but the excited-state eigenvalue
$\Mass$ must be positive.
In fact, it means that we can impose the near-ground-state approximation
$O (\Delta^2) \approx 0$ 
(which was used to obtain Eq. (\ref{eqMcal3})) only if $|\mu| \to 0$.
Then, keeping $\Mass/\mu$ and $\sqrt{|c|}/\mu$ finite, we obtain
\be                                                        \lb{eqMcal3n}
\frac{\Mass}{|\mu|} = -1 - 4 
\frac{a^2 (a^2 -16)}{a^4-64 a - 256} + O (\hbar^2 \Delta^2),
\ \ c<0,
\ee
where $a \equiv \sqrt 2 \sqrt{8 - \sqrt{|c|}/|\mu|}$ and the
values of $c$ and $\mu$
are such that $\Mass$ is positive (which is possible following the
simple numerical estimations).
However, if $|\mu|$ is not small then the total mass is given
by Eq. (\ref{eqMcal2n}) resolved with respect to the
total mass $\Mass$.

Thus, we have ruled out the expressions for
the first-order corrections to the total mass of
a quantum particle with rigidity.
It should be pointed out that
this approach is indeed non-perturbative because all
the formulae above cannot be
obtained by virtue of the perturbation theory assuming small $1/c$
or $\mu$. 
The details and 
other aspects arising at the ``Hamiltonization'' and quantization of the 
theory, 
such as the structure of constraints, spin interpretation
and hidden $SU(2)$ symmetry, can be found  
in Ref. \cite{zlo015}.

\section{Non-minimal particles in dilaton gravity}\lb{s-dg}

In this section we demonstrate how the above-mentioned 
formalism can be also applied to the two-dimensional gravity
because the latter appears to be the special case of
the considered sigma model.
Let us consider the most general action of dilaton gravity 
\begin{equation}
S_{DG}[g, \phi] = \frac{1}{2 k_2^2} \int \drm^2 x \sqrt{-g}
\left[
      D (\phi) R_g  +
      \frac{1}{2} (\partial \phi)^2 +
      V (\phi)
\right] ,                                               \lb{eq3.1}
\end{equation}
where $R$ is the Ricci scalar with respect to metric $g_{\mu\nu}$,
$D$ and $V$ are arbitrary functions. 
By virtue of the Weyl rescaling transformation
$
\bar g_{\mu\nu} = \Omega^2 (\phi)\, g_{\mu\nu}, \ 
\bar\phi = D (\phi),
$
where $\Omega$ is such that
\[
\frac{d D}{d \phi} \frac{d \ln \Omega}{d\phi} = \frac{1}{4},
\]
we obtain the action with only one function of dilaton,
$\bar V (\bar\phi) = V (\phi(\bar\phi))\, \Omega^{-2} (\phi(\bar\phi)) $,
\be
S_{DG}[\bar g, \bar\phi] = \frac{1}{2 k_2^2} \int \drm^2 x \sqrt{-\bar g}
\left[
      \bar\phi R_{\bar g}  +
      \bar V (\bar\phi)
\right] .                                               \lb{eq3.2}
\ee
In the conformal coordinates 
$\bar g_{\mu\nu}=\text{e}^{2\bar u} \eta_{\mu\nu}$
this action can be written in the form  (\ref{eq2.1}) where $N=2$
and
\ba
&&S = k_2^2 S_{DG}, \ \sigma_{a b} 
= \text{offdiag}\, ( -1, -1 ),                              \lb{eq3.3} \\
&& X^a = \{\bar\phi ,\, \bar u \}, \  
U = - \frac{1}{2} \text{e}^{2 \bar u} \bar V (\bar\phi) ,   \lb{eq3.4}
\ea
and therefore we can apply to dilaton-gravitational systems
all the machinery of the previous section.
Namely, knowing the localized solution 
(if it exists, of course) it is supposed to perform the transition
from the Liouville coordinates to the collective coordinates of the 
solution, get rid of zero modes, obtain the effective 
one-dimensional point-particle action with nonminimal corrections,
and then quantize it in a standard way as a mechanical $1D$ system.
In practice, however, it is enough to know the localized 
dilaton-gravity solution and then everything is (almost) automatic
due to the analogy given by Eqs. (\ref{eq3.3}), (\ref{eq3.4}).

Thus, every static solution of the dilaton gravity can be described
in terms of the quantum mechanics of
non-minimal particles provided the
conditions of the formalism above are satisfied.
This opens the  way of the uniform ``Hamiltonization'' and hence
consistent non-perturbative
quantization of any given dilaton
gravity in the vicinity of the static solution, i.e., region 
with non-trivial vacuum.

\section{Supergravity 0-brane solutions as nonminimal particles}\lb{s-sg}

In this section we consider 0-brane solutions of 
the general supergravity in arbitrary spacetime dimensions $D$.
Distinguishing the two cases, dilatonic and non-dilatonic
branes,
we demonstrate on
classical and quantum levels the examples of
how the 0-brane solution of the gauged supergravity 
above can be described
in terms of the one-dimensional
mechanics with rigidity if certain (very natural and evident, such as the
convergence of necessary integrals)
requirements of the field-to-particle transition formalism are satisfied.

\subsection{Dilatonic brane}\lb{s-sg-di}

The supergravity action in the Einstein frame is given by
\ba
&&S = \frac{1}{2 k_D^2}
\int d^D x \, \sqrt{-G^e}
\biggl[
      R_{G^e} - \frac{4}{D-2} (\partial \phi)^2 \nn\\
&& \qquad - 
      \frac{1}{2\cdot 2\ftr} \text{e}^{2 a \phi} F_2^2
\biggr],                                                 \lb{eq4.1}
\ea
where $\phi$ is $D$-dimensional dilaton, $F_2= d A$ is the field
strength of the 1-form potential $A = A_M d x^M$, where
index $M$ runs from $0$ to $D-1$.

Applying the electric-magnetic duality to this action
one can interpret the action of the dilatonic 0-brane as
being magnetically charged under the ($D-2$)-form field
strength $F_{D-2}$:
\ba
&&\tilde S = \frac{1}{2 k_D^2}
\int d^D x \, \sqrt{-G^e}
\biggl[
      R_{G^e} - \frac{4}{D-2} (\partial \phi)^2 \nn\\
&& \qquad - 
      \frac{1}{2\, (D-2)\ftr} \text{e}^{-2 a \phi} F_{D-2}^2
\biggr],                                                 
\ea
which after the Weyl rescaling to the dual frame 
$G^e_{M N} = \text{e}^{-\frac{2 a}{D-3}\phi} G^d_{M N}$
can be rewritten in the form 
\ba
&&\tilde S_d = \frac{1}{2 k_D^2}
\int d^D x \, \sqrt{-G^d}
      \text{e}^{\delta \phi} 
\biggl[
            R_{G^d} + \gamma (\partial \phi)^2   \nn\\
&&\qquad \  -     
      \frac{1}{2\, (D-2)\ftr} F_{D-2}^2       
\biggr],                                             \lb{eq4.3}
\ea
where 
\[
\delta = - a \frac{D-2}{D-3}, \
\gamma = \frac{D-1}{D-2} \delta^2 - \frac{4}{D-2}.
\]
In the dual frame the 0-brane solution is
\ba
&&d s^2_d = H^{\frac{2}{D -3}}
            \left(
            -H^{ - \frac{4}{\Delta}
               } d t^2 +  
            d r^2 + r^2 d \Omega_{D-2}^2
            \right), \\
&& \text{e}^\phi = H^\frac{a (D-2)}{2 \Delta}, \
F_{D-2} = \ast (d H \wedge d t),
\ea
where $H = 1 + (\check\mu / r)^{D-3}$ and 
$\Delta = \frac{a^2}{2} (D-2) + 2 \frac{D-3}{D-2}$.
In the near-horizon region the metric takes the 
$\text{AdS}_2 \times S^{D-2}$ form \cite{bbhs}:
\be
d s^2_d \approx 
- 
\left( \frac{\check\mu}{r} \right)^{2-\frac{4 (D-3)}{\Delta}} d t^2
+ 
\left( \frac{\check\mu}{r} \right)^2 d r^2 + \check\mu^2 d \Omega_{D-2}^2,
\ee
and it is possible to perform the Freund-Rubin compactification
on $S^{D-2}$ of the action (\ref{eq4.3}) to obtain
the following effective gauged supergravity action
\begin{equation}
S = \frac{1}{2 k_2^2} \int \drm^2 x \sqrt{-g}
\text{e}^{\delta\phi}
\left[
      R_g  + \gamma (\partial \phi)^2 +  \Lambda
\right] ,                                               
\end{equation}
where
\[
\Lambda = 
\left(
\frac{D-3}{\check\mu}
\right)^2
\left(
      \frac{D-2}{D-3} - \frac{2}{\Delta}
\right).
\]
After the Weyl rescaling  
$g_{\mu\nu}= \Phi^{-\gamma/\delta^2} \bar g_{\mu\nu}$  
($\Phi=\text{e}^{\delta\phi}$)
it takes the form of the action (\ref{eq3.2}):
\be                                                   \lb{eq4.8}
S = \frac{1}{2 k_2^2} \int \drm^2 x \sqrt{-\bar g}
\left[
      \Phi R_{\bar g}  +
      \Lambda\Phi^{1-\gamma/\delta^2}
\right],                                               
\ee
hence in conformal gauge it also can be written as the special case 
of the
sigma model (\ref{eq2.1}), following Eqs. (\ref{eq3.2}) - (\ref{eq3.4}).
Thus, 0-brane solution of the gauged supergravity 
above can be described
in terms of the quantum mechanics of
non-minimal particles as well.
In turn it means that we can construct the Hamiltonian and 
quantum theories which properly take into account 
the fluctuations of all the acting fields including gravity.

The static solutions of the system described by the
latter action are well-known \cite{lk}
\ba
&& d s^2 = - 
\left[
      \frac{1}{\varrho} 
      (x \sqrt{\Lambda})^{\varrho}
      - 2\sqrt\Lambda M
\right] d T^2  
\nn\\
&& \qquad \ \ \,
+
\left[
      \frac{1}{\varrho} 
      (x \sqrt{\Lambda})^{\varrho}
      - 2\sqrt\Lambda M
\right]^{-1} d x^2, \lb{eq4.9} \\
&&\Phi = \sqrt\Lambda x, \qquad
\varrho \equiv 2 - \frac{\gamma}{\delta^2},              \lb{eq4.10}
\ea
where 
$M$ is the diffeomorphism-invariant parameter
associated with the mass of solution:
\be
M=-\frac{\sqrt{\Lambda}}{2}
\left[
\Lambda (\nabla \Phi)^2 + \frac{1}{\varrho} \Phi^\varrho
\right].
\ee
If the black hole is formed 
then the event horizon appears at 
$
x_H = \Phi_H / \sqrt{\Lambda}
= \frac{1}{\sqrt{\Lambda}}
\left[
      2 \varrho \frac{M}{\sqrt{\Lambda}}
\right]^{1/\varrho},
$
and the surface gravity ($  = 2 \pi T_H$) and entropy are
given by
\be                                                     \lb{eq4.12}
\kappa 
= \frac{1}{\sqrt{\Lambda}}
\left[
      2 \varrho \frac{M}{\sqrt{\Lambda}}
\right]^{1-1/\varrho}, \ \
S = \frac{2\pi}{\kappa_2^2}
\left[
      2 \varrho \frac{M}{\sqrt{\Lambda}}
\right]^{1/\varrho}. 
\ee
Thus, we have all the necessary formulae to study the non-minimal
mechanics of any given dilatonic 0-brane: expressions
(\ref{eq3.2}) - (\ref{eq3.4}) and (\ref{eq4.8}) relate
supergravity with the approach of Sec. \ref{s-ea},
whereas Eqs. (\ref{eq4.9}) and (\ref{eq4.10}) give
the desired solution.
Unfortunately, for the case of general $\gamma/\delta^2$'s 
the metric (\ref{eq4.9}) cannot be rewritten in the conformal
gauge in elementary functions, therefore, each special case requires
a special treatment.
Let us consider the two important examples, $\gamma/\delta^2= 1$
and $\gamma/\delta^2= 0$. 

\bc
{\it CGHS-reduced 0-brane} $\gamma/\delta^2= 1$
\ec

Then we have
\ba
&& \gamma = 4,\ \delta = -2, \ a = 2 \frac{D-3}{D-2}, \\
&& \Lambda = \left(  \frac{D-3}{\check\mu} \right)^2,
\ea
and one can see that
applying the transformation $\Phi= \text{e}^{-2 \phi} $,
$\tilde g_{\mu\nu} = \text{e}^{2 \phi} \bar g_{\mu\nu}$
the action (\ref{eq4.8}) has the form of that of the CGHS
model \cite{cghs}                               
\[
S_{CGHS} = \frac{1}{2 k_2^2} \int \drm^2 x \sqrt{-\tilde g} \, 
\text{e}^{-2\phi}
\left[
      \Phi R_{\tilde g}  +  4 (\partial \phi)^2   +
      \Lambda
\right].                                               
\]                 
Then the static solution of the action (\ref{eq4.8}) at
$\gamma/\delta^2= 1$  is given by
\be                                               \lb{eq4.15}
\bar u_s = C \rho + \bar u_0,\
\Phi_s = \frac{\Lambda}{C^2} \text{e}^{C |\rho| + \bar u_0} + \Phi_0,
\ee
where $\rho = \gamma_l (r-v t)$, $\gamma_l^{-1} = \sqrt{1-v^2}$
and $C<0$, $\bar u_0$, $\Phi_0=2 M$, $v$ are some constants.
We assume the absolute value of $\rho$ in $\Phi_s$
to guarantee that the true
dilaton (which is the logarithm of $\Phi$ up
to a coefficient) remains real.

Before considering the field-to-particle transition
let us clarify the global properties of this solution.
The bar metric $\text{e}^{\bar u_s} \eta_{\mu\nu}$ 
is flat but the true metric,
\be
d s^2 = 
\frac{\text{e}^{C \rho}}{\Phi_s}
(-d t^2 + d r^2)
=
\frac{\text{e}^{C \rho}}{\Phi_s}
(-d \tau^2 + d \rho^2),
\ee
is not, and after the transformation
$
\rho \to R = - \frac{C}{\Lambda}  
\Ln{}{
     \frac{\Lambda}{C^2} \text{e}^{C \rho} + \Phi_0
     }
$
can be written in the Schwarzschild-like form
\be
d s^2 = -
\frac{C^2}{\Lambda}
\left(
      1 - \Phi_0 \text{e}^{\frac{\Lambda}{C} R}
\right) d \tau^2 +
\frac{\Lambda}{C^2}
\left(
      1 - \Phi_0 \text{e}^{\frac{\Lambda}{C} R}
\right)^{-1} d R^2,
\ee
which exhibits the existence at $R_{\text{sing}}= -\infty$ of 
either naked singularity 
or singularity 
under the horizon $R_H = - \frac{C}{\Lambda} \ln{\Phi_0}$ if
$\Phi_0 > 0$.
It is necessary to point out that black holes arising in 
the theories like those described by Eqs. (\ref{eq3.1}), (\ref{eq3.2})
or (\ref{eq4.8}) seem not to be the true black holes because of
the ``semi-completeness'' of the singularity \cite{kkl}.
Namely, all light-like extremals approach a singularity at
an infinite value of a canonical parameter while time- and space-like
ones hit a singularity at a finite value.
Thus, solution (\ref{eq4.15}) describes either naked singularity or
(quasi-)black hole depending on what is the sign of $\Phi_0$
hence if we are interested in solutions with positive $\Phi$
we must suppose the black hole case $\Phi_0 > 0$.
Finally, let us say some words about the value of $C$.
If the black hole is formed and one calculates the surface gravity
from the metric above by virtue of the naive Wick rotation argument 
and compares the result with that obtained from (\ref{eq4.12}) at
$\varrho=1$ then one can conclude that $C=-2/\sqrt{\Lambda}$.
However, below we will keep $C$ unassigned 
for generality.

Let us perform now the field-to-particle transition for this solution
to obtain the action of the type (\ref{eq2.16})
remembering that $X^a = \{ \bar u,\ \Phi \}$ and  signature
of our metric is opposite to that assumed in Sec. \ref{s-ea}.
To obtain the first parameter $\mu$ which is given by Eq. (\ref{eq2.5})
we find that for our solution $L (X^s) =  2 \Lambda \text{e}^{C |\rho|}$
and hence 
\be
\mu =  - \frac{2}{k_2^2} \frac{\Lambda}{C}.
\ee
To find the non-minimal coefficients $\alpha_i$ let us consider
the corresponding Sturm-Liouville eigenvalue problem.
The system (\ref{eq2.20}) takes the form (no summation over $a$)
\be
 - g_a'' (z) - b_a g_a (z) = 0,
\ee
where $z= C \rho$, $b_1 = c_{\hat 1}/C$ and 
$b_2 = c_{\hat 2}/C -1$.
One can see that this system yield free modes and, therefore, no
bound states are formed on the (non-closed and infinite) axis of $\rho$'s.
In fact, the absence of any effective interaction means that the 
particle action does not contain non-minimal 
curvature-dependent terms,
and the CGHS-reduced 0-brane is trivial in this sense.
Another case, $\gamma/\delta^2 =0$, appears to be more
non-trivial in this connection.

\bc
{\it JT-reduced case} $\gamma/\delta^2= 0$
\ec

In this case the bar metric coincides with the physical one,
and the action (\ref{eq3.2}) becomes that of 
Jackiw-Teitelboim gravity
\be
S = \frac{1}{2 k_2^2}  \int d^2 x \sqrt{-g}\, \Phi (R_g + \Lambda)
\ee
whereas the desired static solution is given by
\ba
&&\text{e}^{u_s} = \frac{C}{\Lambda} 
\Cosh{-2}{\sqrt{C/2} \, \rho}, \  
g_{\mu\nu} = \text{e}^{u_s} \eta_{\mu\nu},\\
&& \Phi_s = \Phi_\infty \Tanh{}{\sqrt{C/2} \, |\rho|},
\ea
where $\rho = \gamma_l (r-v t)$, $\gamma_l^{-1} = \sqrt{1-v^2}$
and $C=2 \Lambda^{3/2} M >0$, 
$\Phi_\infty=|\Phi_\infty| \Sign \rho >0$, $v$ are some constants.
We assume the absolute value of $\rho$ in $\Phi_s$
because of the same reason as in the previous case.
Also similarly to the previous case
we find that $L (X^s) =  -2 C \Phi_\infty \Sinh{}{\sqrt{C/2} \, |\rho|}
\Cosh{-3}{\sqrt{C/2} \, |\rho|}$
and hence the first coefficient for the action (\ref{eq2.16}) is
\be                                          \lb{eqJTmu}
\mu = k_2^{-2} \sqrt{2 C}\, \Phi_\infty.
\ee
To find the non-minimal coefficients $\alpha_i$ let us consider
the corresponding Sturm-Liouville eigenvalue problem.
The system (\ref{eq2.20}) takes the form 
\ba
&& - g_1'' (z) - \bigl[ 2 -2 \Tanh{2}{z} + b_1 \bigr] g_1 (z) = 0, \\
&& - g_2'' (z) - \bigl[ 2-6 \Tanh{2}{z} + b_2 \bigr] g_2 (z) = 0, 
\ea
where $z=\sqrt{C/2}\, \rho$ and $b_a = 2 c_{\hat a}/ C$.
Using the asymptotical technique
described in Refs. \cite{zlo015,zlo,zlo006} we can find
that the only solutions of the Sturm-Liouville bound-state problem
are
\ba
&&\{g_1 = \text{const} \Cosh{-1}{z},\, c_{\hat 1} = -C/2 \}, \nn \\
&&\{g_1 = \text{const} \Tanh{}{z},\, c_{\hat 1} = 0 \},\nn \\
&&\{g_2 = \text{const} \Sinh{}{z} \Cosh{-2}{z},\, c_{\hat 2} = 3 C/2 \}, \nn\\
&&\{g_2 = \text{const} \Cosh{-2}{z},\, c_{\hat 2} = 0 \}.  \nn
\ea
The solutions with zero eigenvalues should be omitted since they
do not fulfil the requirements of the field-to-particle
transition approach.
The rest ones are good so the field solution $\{ u_s,\ \Phi_s \}$
can be viewed as the doublet superposition of the two
particles with rigidity:
\ba                                             
&& S_1 = 
- \mu\int \drm s \sqrt{\dot x^2} 
\left(
       1 - \frac{1}{C} k^2 
\right),       \nn\\
&& S_2 = 
- \mu\int \drm s \sqrt{\dot x^2} 
\left(
       1 + \frac{1}{3 C} k^2 
\right),       \nn
\ea
where $\mu$ is given by Eq. (\ref{eqJTmu}).

On the quantum level these particles should be treated as outlined
in Sec. \ref{s-ea-qu}, e.g., the first-order
quantum corrections to their masses are given
for the particle $S_1$ by Eq. (\ref{eqMcal3}) where $c \equiv 3 C/2$
and
for the particle $S_2$ by Eq. (\ref{eqMcal2n}) where $|c| \equiv C/2$
or by Eq. (\ref{eqMcal3n}) provided $\mu$ is small.

\subsection{Non-dilatonic brane}\lb{s-sg-el}

In this case $\phi \equiv 0$ and the action (\ref{eq4.1}) becomes 
singular.
The true effective action for the non-dilatonic supergravity is 
\ba
S = \frac{1}{2 k_D^2}
\int d^D x \, \sqrt{-G}
\biggl[
      R_{G} - 
      \frac{1}{2\cdot 2\ftr} F_{M N} F^{M N}
\biggr],                                                 \lb{eqNDa}
\ea
where $F$ is the field strength of the $U (1)$ gauge field $A_M$.
Let us take the ansatz for the $D$-dimensional metric in the 
form
\be
G_{M N} d x^M d x^N =
g_{\mu\nu} d x^\mu d x^\nu
+ \Expi{-\frac{4 }{D-2} \gaug}
d \Omega^2_{D-2},
\ee
where $\mu,\, \nu = 0,\, 1 $, $g_{\mu\nu}$ and $\gaug$
are functions of $x^\nu$.
If one assumes that the gauge field $A$ is electric then
resolving the Maxwell equation $\nabla_M F^{M N} = 0$
one obtains
\be
F_{t r} = Q \Expi{2 \gaug} \sqrt{-g},
\ee
where $g = \Det{g_{\mu\nu}}$ and $Q$ is the constant related to the
electric charge.
Then the initial action after dimensional reduction on $S^{D-2}$
takes the following form
\ba
&& S = \frac{1}{2 k_2^2}
\int d^2 x \, \sqrt{-g} \Expi{-2\gaug}
\biggl[
      R_{g} - 4\frac{D-3}{D-2} (\partial \gaug)^2 \nn\\
&& \qquad - 
      (D-2) (D-3) \Expi{\frac{4}{D-2}\gaug}
      + \frac{Q^2}{2} \Expi{4\gaug}
\biggr],                                            \lb{eqA2D}     
\ea
i. e., it can be reduced to the form (\ref{eq3.2}) - (\ref{eq3.4}) hence
(being gauged) to the form (\ref{eq2.1})
that means that its localized
static solutions can be interpreted
as nonminimal particles as well
(however,   some subtle thing exists here so let us demonstrate 
the approach step by step).
The corresponding equations of motion are
\ba
&&R_g + \frac{D-3}{D-2}
\left[
     \frac{(\partial \Phi)^2}{\Phi}
     - 2 \nabla_\nu 
     \left(
            \frac{\partial^\nu \Phi }{\Phi}
     \right)
\right] -                                         \nn\\
&& \qquad \qquad
 (D-3) (D-4) \Phi^{\frac{\delta}{D-2}}
- \frac{Q^2}{2} \Phi^\delta = 0,                   \lb{eqEM11}  \\
&&
\nabla_\mu \nabla_\nu \Phi +
\frac{D-3}{D-2}
\left[
      \frac{g_{\mu\nu}}{2} \frac{(\partial \Phi)^2}{\Phi}
      - \frac{\partial_\mu \Phi \, \partial_\nu \Phi}{\Phi}
\right] + \nn\\
&& \qquad  \qquad
\frac{g_{\mu\nu}}{2}
\left[
      (D-2) (D-3) \Phi^{\frac{D-4}{D-2}} - \frac{Q^2}{2 \Phi}
\right]=0,                                          \lb{eqEM12}
\ea
where $\Phi = \Expi{\delta\gaug}$, $\delta \equiv -2$ and
$\nabla$ means the covariant derivative with respect to 
the metric $g$.

In this paper we are mainly interesting in 0-branes, moreover,
in the near-horizon region.
The $D$-dimensional 0-brane solution is given by
\ba
&&d s^2 = 
- H^{-2} 
d t^2 +
H^{\frac{2}{D-3}}
\left(
       d r^2 + r^2 d \Omega^2_{D-2}
\right), \nn\\
&&A_t = -H^{-1},
\ea
where $H = 1+ (\check\mu/r)^{D-3}$.
In the near-horizon region spacetime becomes
$\text{AdS}_2 \times S^{D-2}$ and we have
\ba
&&d s^2 = 
- 
\left(
      \frac{r}{\check\mu} 
\right)^{2 (D-3)}
d t^2 +
\left(
      \frac{r}{\check\mu} 
\right)^{-2}
d r^2
+
\check\mu^2 d \Omega^2_{D-2}, \nn\\
&&A_t = - (r / \check\mu)^{D-3},
\ea
whereas $\gaug$ becomes constant.

Therefore, we will seek for those static
solutions of Eqs. (\ref{eqEM11}),
(\ref{eqEM12}) whose spacetime part has the desired
$\text{AdS}_2 \times S^{D-2}$ form.
One can check that the following pair,
\ba
&& \Expi{u_s} = \frac{C}{\Lambda} \Cosh{-2}{\sqrt{C/2}\rho}, \ \;
g_{\mu\nu} = \Expi{u_s} \eta_{\mu\nu},        \lb{eqEM21}\\
&& \Expi{-2 \gaug_s} \equiv \Phi_s =
\left[
      \frac{\Lambda/2}{(D-3)^2}
\right]^{1- D/2},                             \lb{eqEM22}
\ea
where 
$\Lambda=  2 (D-3)^2
\left[
      \frac{(D-2) (D-3)}{Q^2/2}
\right]^{\frac{1}{D-3}}$, $C$ is a positive constant
and $\rho$ is the same as in previous sections, 
appears to be the solution of Eqs. (\ref{eqEM11}),
(\ref{eqEM12}), besides the gravitational part has the desired 
AdS form $R_{g_s} = \Lambda $.

Further, performing the Weyl rescaling 
$\{ g ,\, \gaug \} \to \{ \bar g, \, \Phi \}$  such that
$g_{\mu\nu} = \Phi^{-\gamma/\delta^2} \bar g_{\mu\nu}$
we can rewrite the action (\ref{eqA2D}):
\be
S = \frac{1}{2 k_2^2}
\int d^2 x \sqrt{-\bar g} \, \Phi
\left[
       R_{\bar g} + \bar\Lambda (\Phi)
\right],
\ee
where $\delta = -2$, $\gamma/\delta^2 = - \frac{D-3}{D-2}$, and
\[
\bar\Lambda (\Phi) \equiv
- \Phi^\frac{D-3}{D-2}
\left[
      (D-2) (D-3) \Phi^{\frac{\delta}{D-2}} -
      \frac{Q^2}{2} \Phi^\delta
\right],
\]
therefore, in the Liouville gauge 
$\bar g_{\mu\nu} = \Expi{\bar u_s} \eta_{\mu\nu}$
the Weyl-transformed action obtains the form (\ref{eq2.1}) where  
\ba
&&\rphi^1 \equiv 
\Expi{\bar u_s} =
\frac{C/2}{D-3} \Phi^\frac{D}{D-2}_s 
\Cosh{-2}{\sqrt{C/2}\rho} , \nn\\
&&\rphi^2 \equiv  \Phi_s, \ \
U \equiv \bar\Lambda (\rphi^2)\, \rphi^2 \Expi{\rphi^1}. \nn
\ea
However, it turns out that the effective (sigma-model) Lagrangian
vanishes on this doublet solution,
therefore, the non-dilatonic brane in the near-horizon
region cannot be described by the
two-component sigma model.
In fact, it is caused by the fact that the electric component
of this solution happens to be constant in this region hence
the whole solution can not be regarded as two-component.

To construct the appropriate one-component effective action
let us fix $\Phi=\Phi_s$ rather than $g$.
Then in the Liouville gauge the remaining equation
(\ref{eqEM11}) can be 
written as the Liouville equation
\be
\eta^{\mu\nu} \partial_\mu \partial_\nu u - \Lambda \Expi u  = 0,
\ee
hence the appropriate one-component effective theory
is the Liouville one.
The field-to-particle transition for the Liouville
theory was previously considered
in Ref. \cite{zlo006} so let us use the results obtained there.
Defining
\[
\beta \equiv 1/2, \
m = 2 \Lambda, \
\zeta \equiv m C = 2 \Lambda C,  \ S \to 2 k_2^2 S,
\]
we straightforwardly obtain the desired non-minimal particle
action
\be
S_{\text{eff}} = - \mu \int d s \sqrt{\dot x^2}
\left(
       1 - \frac{k^2}{\Lambda C}
\right),
\ee
where $\mu = 2 k_2^{-2} \sqrt{2\Lambda C}$.
Finally, also we immediately 
obtain that the first-order quantum correction 
to the bare mass $\mu$ is given 
by Eq. (\ref{eqMcal2n}) 
(or by Eq. (\ref{eqMcal3n}) if $\mu \to 0$)
where $|c| \equiv \Lambda C/2$.

\section{Discussion}\lb{s-d}

Let us enumerate the main aims achieved in the previous sections.
First, in Sec. \ref{s-ea}
the field-to-particle transition formalism was generalized
for the sigma model.
It was shown that the field fluctuations cause  the
appearance of the nonminimal curvature-dependent terms
in a final effective action.
Using it we performed the quantization of the sigma model
in the vicinity of the nontrivial vacuum induced by a certain
soliton (solitary-wave) solution of the model.
Then in Sec. \ref{s-dg} it was demonstrated 
how one can directly apply the approach of Sec. \ref{s-ea} 
to the  two-dimensional dilaton gravity in the Liouville gauge.
Then the black hole solution which is the
doublet consisting of graviton and dilaton components
can be effectively interpreted as a massive point-particle
with rigidity (curvature).
Further, in Sec. \ref{s-sg} the  field-to-particle transition
approach was applied to the supergravity 0-brane solutions in
the near-horizon approximation.
Also as in previous cases the final effective action appeared  
belonging to the class of non-minimal point-particle actions,
and again it has opened the way of uniform study of supergravity
solutions.

Completing the summarizing of
the part one we would like also to outline something that
goes behind the field-to-particle transition and field theory itself.
Namely, the fact that the physically admissible
solutions of so different field theories as the sigma model,
dilaton and supergravity (which themselves are the reductions of
more higher dimensional theories) eventually can be reduced to 
the description 
of the same object, non-minimal particle, can be read also
``from right to left''.
It may happen that it is not a mere coincidence: the 
point particles can be not only the
end product but also the underlying base of modern high-energy
theory even more fundamental than the strings.
Indeed, being due to pointness geometrically simpler than a string 
the non-minimal particle nevertheless has the rich mathematical 
structure due to self-interaction \cite{txt-nm}.
Then the hierarchy of this world would become even more natural:
{\it
high-dimensional point particles are the base,
strings can be regarded as the path trajectories of these particles},
membranes (1-branes)
are surfaces swept by strings, etc.;
different types of strings are caused by existence of
several paths of the
same particle;
other (non-fundamental) ``point'' particles are excited states 
of the string(s),
and so on.
The holographic principle holds for the field-to-particle
(string-to-particle) point of view as well: 
the string which has 
two-dimensional worldsheet space is described by the (non-minimal)
particle whose worldvolume spacetime is one-dimensional.

However, what is the concrete mechanism of how a 
point particle could yield
the extended object like string or brane?
In the part two we try to answer this question.

\section{Part two:
The ``microscopical'' theory of
strings and branes
}\lb{s-ps}

When discussing the results of Secs. \ref{s-ea} - \ref{s-sg}
in the previous section it was conjectured that the hierarchy 
of this world ``begins and ends'' with the point particle.
The aim of this section is to deepen and improve this conjecture
by virtue of proposing of a concrete mechanism.

\subsection{Worldvolume holography: Transition from point particle
to extended object via fuzzy state}\lb{s-ps-ga}

Let us begin with the simplest action of the 
bosonic point particle
of mass $\mu$
in the $D$-dimensional spacetime 
\be                                                \lb{eqPwR}
S_p = -\mu \int d s \sqrt{-\dot X^\nu \dot X_\nu},
\ee
or in the root-free form
\be                                                \lb{eqPnoR}
S_p = \frac{1}{2} \int d s 
\left(
 \eta^{-1} \dot X^\nu \dot X_\nu -\eta \mu^2 
\right),
\ee
where $\eta$ is the ``einbein'' - the auxiliary non-propagating
field related to the one-dimensional metric
of the particle worldline (we follow the notations of
the Polchinski's book \cite{pboo}).

Let us try to understand this action in the more general way.
If one introduces the operator of the transport along a 
worldline - vector ${\bf T}$ such
that 
\be                                                \lb{eqTV0}
{\bf T}^{(0)} =  \PDer{}{s}, \ \ \text{i.\,e.,} \ \
T^{(0)}=
\left(
\begin{array}{l}
\  1      \\
\left.\begin{array}{l}
0      \\
\, \vdots\\
0     \\
\end{array}\right\} D-1 
\end{array}
\right),
\ee
where the components of the column are taken on the space of 
the Frenet vectors corresponding to the particle's worldline:
${\bf T}^{(0)} = T^{(0) \alpha}\, \partial_{\sigma^\alpha}
= (T^{(0)\, \alpha})_\nu d X^\nu$
where $\sigma^i$, $i=1,...,D-1$, are the 
appropriate coordinates toward the 
directions orthogonal
to the worldline.
Then the action (\ref{eqPnoR}) can be equivalently rewritten as
\be                                                \lb{eqPnoRT}
S_p = \frac{1}{2} \int d s 
\left[
 \eta^{-1}\, \text{tr} 
\left\{
{\bf T}^{(0)} X^\nu \, {\bf T}^{(0)} X_\nu 
\right\}
-\eta \mu^2 
\right].
\ee
The
quantum uncertainty (the interaction between a microworld object 
and a macroscopical device governed by $\hbar$) 
and perhaps
the classical chaos (the secular dynamical instability
which leads to the uncertainty of trajectory at large $\Delta s = s-s_0$
through
the so-called ``forgetting of initial data'') 
may  
affect the transport vector and break the initial structure  
(\ref{eqTV0}) down,
\[
{\bf T} = U \, {\bf T}^{(0)} U^{-1},
\]
so
in general case the transport is described by:
\be
{\bf T}^{(0)} \to                                                \lb{eqTVt}
{\bf T} = T^\alpha \partial_{\sigma^\alpha},
\ \ 
T =
\left(
\begin{array}{l}
T^0      \\
T^1      \\
\, \vdots\\
T^{D-1}     
\end{array}
\right),
\ee
where $\sigma^0 \equiv s$, and
$T^1, ... ,T^{D-1}$ may happen to be non-zero
(moreover, from now we will regard $T$'s as the matrices
whose properties become clear below - after Eq. (\ref{eqMB})), and
\be                                             \lb{eqDim}
\text{dim} \{\sigma\} \leq 
\text{maxdim} \{\sigma\} =
\text{dim} \{X\} = D. 
\ee
Replacing in Eq. (\ref{eqPnoRT}) ${\bf T}^{(0)}$ 
by ${\bf T}$ from Eq. (\ref{eqTVt}) we obtain 
the action of what we will call
the {\it fuzzy (matrix) particle}
\be                                                \lb{eqFP}
S_{\text{seed}} =- \frac{1}{2 l_p} \int d s \,
 \eta^{-1}
\left\{
T^\alpha,\, T^\beta 
\right\}
\partial_\alpha X^\nu \partial_\beta X_\nu,
\ee
where it was imposed $\mu \equiv 0$
(in fact, there was no need in $\mu$ {\it ab initio} but we had to
use it to make the considerations smooth),
and of course
$\partial_\alpha = \partial_{\sigma^\alpha}$.
Besides, we have introduced the characteristic length of 
particle's path $l_p$ as a normalizing overall coefficient
assuming instead the dimensionlessness of $\eta$.

The sense of the subscript at $S$ will become clear below -
the action we have obtained will appear to be the seed action
for some forthcoming object.
Indeed, the action of the latter can be 
obtained by the averaging of the seed
action over all the deviations 
$\{\sigma^1,\, ...,\, \sigma^{D-1}\}$ - the averaged picture is
the inevitable result of any procedure of measurement.
So, 
\ba                                                \lb{eqMB}
&&S_{MB} = 
\frac{
      \int d \sigma^1 ... \, d \sigma^{D-1} \sqrt{-\gamma_{1\by D-1}}
      \, S_{\text{seed}}
     }
     {
      \int d \sigma^1 ...\, d \sigma^{D-1} \sqrt{-\gamma_{1\by D-1}}
     }      \nn\\
&&\qquad
=
-\frac{1}{V_{D}} \int 
[d \sigma] \sqrt{-\gamma}
\, \gamma^{\alpha\beta}
\partial_\alpha X^\nu \partial_\beta X_\nu,
\ea
where we have defined $[d \sigma] = d \sigma^0 ... \, d \sigma^{D-1}$,
the total volume $V_D \equiv V_{0\by D-1} = l_p V_{1\by D-1}$, besides
$\gamma_{1\by D-1} \equiv \text{det}\, \frac{1}{2}\{T^i,\, T^k\}$ 
$(i,k = 1,\, ..., D-1)$ is a function of $\sigma^i$'s, such that
we obtain the effective  worldsheet metric in the vielbein form
$
\gamma^{\alpha\beta} \sqrt{-\gamma} 
=
\frac{1}{2}
\{T^\alpha, T^\beta\} \sqrt{-\gamma_{1\by D-1}/\eta^2 },
$
or, shortly,
\be                                            \lb{eqG}
\gamma^{\alpha\beta}  
=  \frac{1}{2}
\{T^\alpha, T^\beta\},
\ee
hence $T^\alpha$'s coincide 
with the gamma matrices generalized for
curved spacetime \cite{foc} 
(for the most general expression for $T$'s see Ref. \cite{txt-dil})
and hence realize the Clifford algebra 
for the vielbein spacetime for which
the Hilbert space forms representation.
In general case $\gamma^{\alpha\beta}$ is not necessary a 
symmetric matrix - it can have an antisymmetric part if $\partial X$'s
do not commutate (for different reasons like the non-commutative
geometry \cite{fgr} or due to their 
contraction with an antisymmetric tensor over
$\mu\nu$ indices); in view
of aforesaid, the antisymmetric part
is given by the (generalized) Dirac commutator 
$\Sigma^{\alpha\beta} =\frac{1}{2} [\gamma^\alpha,\gamma^\beta]$.

Be that as it may be, 
keeping in mind Eq. (\ref{eqDim}) we see that $S_{MB}$ describes
the embedding (or the superposition of embeddings) of the total
dimension $\text{maxdim} \{\sigma\} \leq D$, and we will call it (them) 
the {\it metabrane}.
Thus, the worldsheet metric of the metabrane 
is induced and governed by the averaged set of all
the possible path deviations of a quantum point particle.
Besides, the number of metabrane's worldsheet dimensions can
vary along with evolution but becomes fixed 
once the averaging
(measurement) is performed, as in Eq. (\ref{eqMB}).

The proposed mechanism 
(\ref{eqPnoR})$\to$(\ref{eqFP})$\to$(\ref{eqMB}) 
should not be confused
with the ideology of the matrix theory \cite{bfss}: 
despite that in the both cases 
the point-particle quantum
mechanical paradigm is exploited there exists the huge difference
between how these approaches act.
Namely, matrix theory is 
the eleven-dimensional SYM QM possessing the symmetry $U (N)$
which does not imply any ``smearing'' of the
transport operator along the single-particle worldline. 
Below it will be shown that there is nothing common also 
in the aims and final results.

Before we proceed to the special cases some conclusive
comments are in order.
The fact that we have obtained from the point particle 
(\ref{eqPwR}) the extended
object (\ref{eqMB}) just taking into account the 
uncertainty can be understood
in the following way.
Everyone intuitively understands that
the real, i.e., non-bare, particle in $D$-dimensional spacetime
is observed ($=$ measured) as the
$(D-1)$-dimensional
``cloud'' consisting of the particle's virtual paths 
and hence
nothing prevents them from deviating from the classical
(averaged) trajectory governed by ${\bf T}^{(0)}$.
Therefore, the real particle can be no more regarded as a point object.
However, it cannot be regarded also as a (deterministic) extended
object unless we average over all the possible
deviations.
Once we have performed the averaging we have obtained the 
extended object which can be considered ``as is'', i.e.,
independently of the underlying
structure because 
it does possess
the new set of the (collective) degrees of freedom.

\subsection{Special cases of metabrane: strings plus branes plus
worldsheet fields}\lb{s-ps-sc}

The special cases of the split of the metabrane
(\ref{eqMB}) are determined by the concrete features of how
the transport vector (matrix, more correctly) is ``smeared''.
Let us glance over the most instructive of them.

{\it Single string}.
This case in some sense the first-order (averaged) correction to the bare
point particle.
If one supposes that uncertainty causes the transport
toward the only one extra orthogonal 
direction, for definiteness, $\sigma^1$
in addition to the $\sigma^0 = s$ one such that the
deviation-induced metric $\gamma$ takes the form
\be                                                     \lb{eqGss}
\gamma^{\alpha\beta} =
\left(
\begin{array}{ccc}
\tilde\gamma^{0 0} & \tilde \gamma^{0 1} &     \\
\tilde\gamma^{1 0} & \tilde\gamma^{1 1} &     \\
  &   &   \ddots
\end{array}
\right),
\ee
rest components are zeros, $\tilde\gamma^{a b}$ and $X^a$ are functions
of $\sigma^a$'s ($a,b=0,\,1$),
then the metabrane (\ref{eqMB}) after integrating other dimensions
out reduces to
\ba                                                \lb{eqSS}
S_{SS} = 
-\frac{V_{2\by D-1}}{V_{0\by D-1}} \int 
d \sigma^0 d \sigma^1
\sqrt{-\tilde\gamma}
\, \tilde\gamma^{a b}
\partial_a X^\nu \partial_b X_\nu.
\ea
This is almost the action of the 
(Brink-Di Vecchia-Howe-Deser-Zumino-) Polyakov string \cite{pol}, the only
thing which yet remains unclear is the overall constant.
If one requires that the space 
$\{ \sigma^0, \sigma^1, ..., \sigma^{D-1} \}$ has the following
topology:
$Topology \{ \sigma^0, \sigma^1, ..., \sigma^{D-1} \}
=
Topology \{ cylinder \}$ $\times$ $
Topology \{ \sigma^2, \sigma^3, ..., \sigma^{D-1} \}
$, i.e., 
$
S^1 \times R^1 \times
Topology \{ \sigma^2, \sigma^3, ..., \sigma^{D-1} \}
$
then 
$V_{0\by D-1}= 2 \pi R_{\text{charact}} l_p V_{2\by D-1}$,
and we exactly obtain the Polyakov action
\ba                                                \lb{eqSSP}
S_{SS} = 
-\frac{1}{4\pi\alpha'} \int 
d \sigma^0 d \sigma^1
\sqrt{-\tilde\gamma}
\, \tilde\gamma^{a b}
\partial_a X^\nu \partial_b X_\nu,
\ea
where the constant
$\alpha' = R_{\text{charact}} l_p/2$ is the Regge slope.
One may wonder why so much attention was paid to the overall coefficient
which is not very important 
for the classical dynamics of the string in the sense it
does not appear in equations of motion?
There are the two reasons we were led by.
At first, we demonstrate how the global properties of an embedding
generate the characteristic constant of this embedding.
Second, such coefficients are overall only in the case of the 
single string whereas in the cases when we have a number of
embeddings they may play much more important role of either
weight or coupling constants.
At least, they determine the characteristic length scale of an
embedding: whether it belongs to the microscopical world or
(running ahead) to the cosmological scale.

{\it Multiple strings: metabrane as ``storage'' place}.
Let us suppose now that the uncertainty affects all the components
of the transport vector (matrix) but in such a way that $\gamma$ can be
decomposed into the direct product of $2\times 2$ matrices of 
Lorentzian signature 
$\gamma_{i\by k}^{a b} = \gamma_{i\by k}^{a b} (\sigma^i, \sigma^k)$,
$i \not= k$.
Correspondingly, the
metabrane becomes split into the superposition of strings
(or strings and particles, in dependence on the structure and on
whether $D$ is odd or even).
For the block-diagonal case,
\be
\gamma^{\alpha\beta} =
\left(
\begin{array}{ccc}
\gamma_{0\by 1}^{a b} &                       &     \\
                      & \gamma_{2\by 3}^{a b} &     \\
  &   &   \ddots
\end{array}
\right),
\ee
we have from Eq. (\ref{eqMB})  the superposition of $D/2$
strings ($D$ is assumed even):
\ba
&&S_{MB=\Sigma S} = - \Sum{i=0}{\frac{D}{2}-1}
\alpha_{2 i\by 2 i+1}  
\int d\sigma^{2 i} d\sigma^{2 i+1} 
\sqrt{-\gamma_{2 i\by 2 i+1}} \nn\\
&& \qquad \qquad \ 
\times
\gamma^{a b}_{2 i\by 2 i+1}
\partial_a x^\nu_{(i)} \partial_b x_{(i)\nu},
\ \ \ ^a_b = 2 i,\, 2 i+ 1, 
\ea
where $x_{(i)}$ is the position of the $(i+1)$th string, and
it was imposed that
the metabrane's $X$ is decomposed into the sum:
\ba                                            
&&X^\nu (\sigma^\alpha) = 
x_{(0)}^\nu (\sigma^0,\, \sigma^1) +
x_{(1)}^\nu (\sigma^2,\, \sigma^3) + ... \nn\\
&&\qquad \qquad \
+ x_{\left(\frac{D}{2}-1\right)}^\nu 
(\sigma^{D-2},\, \sigma^{D-1}).                           \lb{eqXSms}
\ea
The values of the weight constants $\alpha$  are determined again
after integrating appropriate extra dimensions out
by the topology of a split submanifold and its characteristic size:
\be                                                       \lb{eqWC}
\alpha_{0\by 1} = 
\frac{V_{2\by 3 \by ... \by D-1}}{V_{0 \by ... \by D-1}}, \
\alpha_{2\by 3} = 
\frac{V_{0\by 1 \by 4 \by ... \by D-1}}{V_{0 \by ... \by D-1}}, \
\text{etc.},
\ee
that is, they are, in fact, inverse volumes of embeddings.

Some comments of how all this can be used should be done.
Recently it is believed that the five consistent superstring theories
due to the mutual dualities are the part of some underlying M-theory.
On the other hand,
from the viewpoint of the proposed approach the metabrane plays
the role of an absorbing theory because the strings 
(and, as will be shown below, branes)
appears to be parts of it. 
If there exists the temptation to proclaim metabrane as
the desired M-theory but it is not so: the $M$ theory in its 
narrow sense \cite{mth1,mth2} implies the 
$11D$ Poincar\'e-invariant theory whose decoupling limit
yields the $10D$ closed oriented superstring IIA,
whereas the metabrane itself is much less tricky: 
it is rather the ``storage'' place of 
several $\ast$-branes including strings; one may assume,
however, that the string-string
dualities can be regarded as the relations between the components 
of the metabrane.

Further, 
in its turn the metabrane also has the underlying theory - 
the microscopical
theory of a fuzzy particle from which 
it was obtained by virtue of the averaging
(definitely, 
there appears some mishmash with the multiple use of
the word ``microscopical'' - strings themselves are  
microscopical objects).

{\it Worldsheet fields}.
The good question is how the worldsheet fields can arise
from the metabrane's internal structure.
It was mentioned above that dilaton can be included into the
structure of transport matrix \cite{txt-dil}
but what is about other fields?
To answer it let us consider the appearance of the worldsheet curvature
for the case of the single string:
assume the structure (\ref{eqGss}) but 
$X$  is now complex (of course, in all the actions above the second $X_\nu$
is assumed to be replaced by its conjugation $X^\ast_\nu$):
\be                                            
X^\nu (\sigma^\alpha) = 
x^\nu (\sigma^0,\, \sigma^1) +
i\, R^\nu (\sigma^0,\, \sigma^1),                          \lb{eqXSwf}
\ee
where $x= \Re X$ is the position of the string, $R=\Im X$ appear to
be additional degrees of freedom.
The Eq. (\ref{eqMB}) 
after integrating other dimensions
out yields
\ba                                                     
&&S_{SS} = 
-\frac{V_{2\by D-1}}{V_{0\by D-1}} \int 
d \sigma^0 d \sigma^1
\sqrt{-\tilde\gamma}
\, \tilde\gamma^{a b}
\bigl(
\partial_a x^\nu \partial_b x_\nu               \nn\\
&& \qquad \quad \;  +
\partial_a R^\nu \partial_b R_\nu
\bigr),
\lb{eqSSwf}
\ea
and the latter term can be associated with the induced 
worldsheet curvature,
\be                                                  \lb{eqIR}
\frac{V_{2\by D-1}}{V_{0\by D-1}}
\partial_a R^\nu \partial_b R_\nu \equiv
\frac{\lambda}{4\pi} R_{a b},
\ee
where $R_{ab}$ is the Ricci tensor constructed from $\tilde\gamma_{a b}$,
so that we eventually have the string with the induced worldsheet
gravity
\ba                                                
&&S_{S} = 
-
\int 
d \sigma^0 d \sigma^1
\sqrt{-\tilde\gamma}
\biggl(
\frac{1}{4\pi\alpha'} 
\tilde\gamma^{a b}
\partial_a x^\nu \partial_b x_\nu            \nn\\
&& \qquad \quad \;  +
\frac{\lambda}{4\pi} R
\biggr), \ \ ^a_b = 0,\,1.                                      \lb{eqSSP2}
\ea
The further generalizations can be done
by analogy exploiting the decompositions (\ref{eqXSms})
and (\ref{eqXSwf}).
One may ask, however, is the capacity of the combination 
$\partial_a R^\nu \partial_b R_\nu$  enough to store
all the information about the curvature?
The answer is ``yes'' that can be confirmed by simple calculation
of independent components.

{\it All together: the string-brane world}.
In general case, the metabrane can contain in addition to strings
the branes of several worldsheet dimensions
(throughout the paper we understand the branes
in the most broad sense of Ref. \cite{hlp} 
besides of the black hole p-branes \cite{hs}
and superstring D-brane solutions \cite{dbrane1,dbrane2}).
The 3-branes are of special interest because of the anthropic
principle and not so far proposed
(Antoniadis-Arkani-Hamed-Dimopoulos-Dvali-Gogberashvili-)Randall-Sundrum 
cosmological
scenario(s) \cite{bw1,bw2,bw3,z0007075}.
The 3-brane can be separated when: (a)
the metabrane's worldsheet metric can be decomposed into the form 
``$4\times 4$ metric times the rest'',
and (b) $X$ is split into the sum
\be                                           
X^\nu (\sigma^\alpha) = 
x^\nu (\sigma^0, ..., \sigma^3) +
i\, \sum\limits_f R^\nu_{(f)} (\sigma^0, ..., \sigma^3) 
+ \, \{\text{rest}\},                          \lb{eqXStb}
\ee
where $x$ is the position of the 3-brane,
$R_{(f)}$'s are ``square roots'' of the worldsheet fields
including the confined 
3-brane Standard Model 
(especially if one assumes the more realistic
supersymmetric case described in the next section).
The term
``rest'' means all other degrees of freedom corresponding
to strings 
(by analogy with Eq. (\ref{eqXSms})), 
worldsheet fields 
(by analogy with Eq. (\ref{eqXSwf})) 
and their possible
interactions between themselves and with 3-brane.
Then the metabrane becomes
the following superposition
\ba                                                        
&&S_{MB} = S_{\text{3-brane}} + \Sigma S_{\text{strings}}
+ S_{\text{worldsheet fields}} \nn\\
&& \qquad\quad \ \
+ S_{\text{mutual interactions}},                          \lb{eqSBW}
\ea
hence one can see that the metabrane does contain both the microscopical
embeddings (strings) and cosmological-scale ones (such as 
our 3-brane Universe): their characteristic 
sizes are encoded in the corresponding weight constants 
$\alpha_{i_1\by i_2 \by ... \by i_k} $ which
actually are the inverse volumes of the embeddings,
like those in Eq. (\ref{eqWC}).
When the 3-brane universe is small (the ``early universe'') 
the value of its weight constant is comparable with those
of strings and decreases as the 3-brane expands.

\begin{figure}[htbp]
  \begin{center}
    \leavevmode
    \epsfxsize=0.9\columnwidth
    \epsfbox{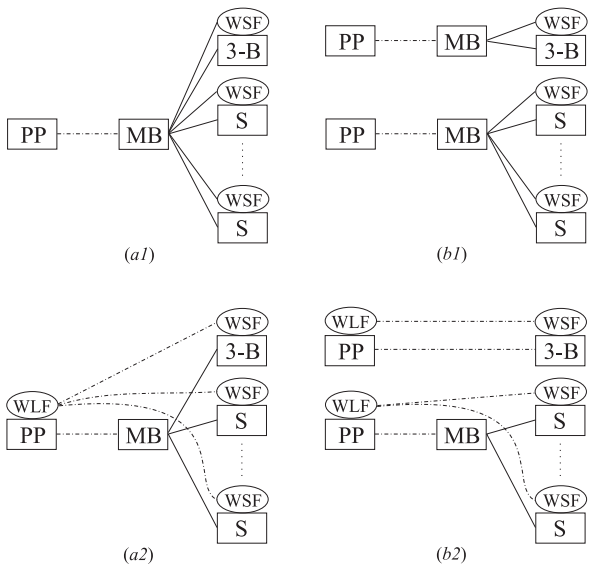}\vskip1mm
\caption{Some of the most characteristic ways 
of string-brane world formation.
The dash-dotted line means subsequent applying of particle-to-fuzzy-particle
``smearing'' and fuzzy-particle-to-metabrane averaging procedures,
the solid line means the way of how the metabrane's worldsheet metric
is decomposed; ``PP'', ``MB'', ``3-B'', ``S'' mean respectively the
point particle, metabrane, 3-brane, string, and ``WLF''
and ``WSF'' mean the worldline and worldsheet fields.
}
    \label{f-pp}
  \end{center}
\end{figure}

Finally, we must say about the varieties of the formation of the
string-brane world in addition to the simplest case described above, see
Fig. \ref{f-pp} ({\sl a1}).
Namely, despite the transition from particle to metabrane is more
or less unique due to the tough condition (\ref{eqG}) or \cite{txt-dil}
for the transport matrix but:

(i) it is unclear how many fundamental particles we have: 
for example,
in the column [({\sl b1}),({\sl b2})] of subfigures of Fig. \ref{f-pp} 
everything starts with two point particles:
one particle yields strings whereas another one produces a
3-brane unlike the column [({\sl a1}),({\sl a2})];
this question is strongly related with the total number of dimensions $D$:
in the case [({\sl a1}),({\sl a2})] the five strings and 3-brane can be 
incorporated into the metabrane if its worldvolume dimension is not
less than $4+5\times2 = 14$ that in turn assumes $D\geq 14$ whereas
in the case [({\sl b1}),({\sl b2})]  we have the two metabranes whose
worldvolume dimensions are respectively $4$ and $5\times 2=10$ 
(for five strings) so $D$ must be not less than ten;

(ii)
the mechanism of producing of worldsheet fields is non-unique:
for example,
in the row [({\sl a1}),({\sl b1})] 
the particle yields the metabrane
which in turn produces both the desired
$\ast$-branes and worldsheet fields on them
but nothing hinders us from assuming the initial particle
having the worldline field which would directly induce the worldsheet ones
as drawn in the row [({\sl a2}),({\sl b2})]
rather than one uses the constructions like (\ref{eqXSwf});

(iii) the 3-brane and string are required by anthropic principle and 
gauge symmetry but yet there exists a lot of
several ways of the split of the metabrane (\ref{eqMB}) (or,
more precisely, (\ref{eqsuMB}))
into the $\ast$-branes
so it is possible that 0-branes and $(p \not= 3)$-branes 
do appear as well.

\noindent

\subsection{Including fermions: supermetabrane}\lb{s-ps-fe}

The scheme given in the two previous subsections  
is quite general but some of its
components need to be corrected.
Namely, everything said before holds but
the strings must be supersymmetric rather than bosonic because
the latter (a) do have tachyonic states, (b) do not have fermions,
and the best way to incorporate the fermions into the picture is
supersymmetry.
Also, it is desirable to regard the 3-brane as the supersymmetric as well:
then the bosonic coordinates are where we live in
whereas the
fermionic part yields the Fermi sector of the confined Standard Model.

So, below we will try to perform the 
particle-to-fuzzy-particle-to-metabrane 
transition for the supersymmetric point particle in $D$ dimensions.
Let us consider the 
one-dimensional (worldline) $N=1$ supergravity \cite{spart}
given by the
following superspace action of the massless superparticle 
\be                                                   \lb{eqsuP}
S_p = \frac{i}{2}\int d \theta d s\, \Lambda^{-1} 
\nabla_\theta \bar Y^\nu \nabla_s Y_\nu,
\ee
where $\theta$ is the Grassmann coordinate,
there is no spin connection hence the spinor covariant
derivatives are
\be
\nabla_\theta = \partial_\theta + i \theta \partial_s, \ \
\nabla_s = \partial_s,
\ee
and the superfields $Y$ and $\Lambda$ are the supersymmetric
counterparts of
the particle's position vector $X$ and einbein $\eta$, respectively,
\ba
&& Y^\mu = X^\mu + i \theta \psi^\mu, \\
&& \Lambda = \eta + i \theta \chi.
\ea
In the same manner as in Sec. \ref{s-ps-ga} we assume that the uncertainty
causes the transport also
toward the directions 
$\sigma^i$, $i=1,...,\text{maxdim}\{\sigma\} -1$, 
orthogonal to the
worldline:
\be                                                         \lb{eqsuTVt}
{\bf T}^{(0)} =
\partial_s  
\to                                                
{\bf T} = \gamma^\alpha \partial_{\alpha},
\ee
where we have taken into account our knowledge about the forthcoming
Eq. (\ref{eqG})
and have neglected Ref. \cite{txt-dil} for simplicity.
Also for simplicity we assume that uncertainty does not cause
the transport toward extra Grassmann directions because eventually
it would require the changes of the supersymmetry index $N$.

Substituting all these formulae back into the initial 
action (\ref{eqsuP}) and integrating over the Grassmann coordinate
we eventually obtain the action of the fuzzy superparticle
(cf. Eq. (\ref{eqFP}))
\ba
&&S_{\text{seed}} = - \frac{1}{l_p}
\int \frac{d s}{\eta}
\biggl[
\frac{1}{2}
\{\gamma^\alpha, \gamma^\beta \}
\partial_\alpha X^\nu \partial_\beta X_\nu    \nn\\
&& \qquad \qquad
+ 2 i \bar\psi^\nu \gamma^\alpha \partial_\alpha \psi_\nu
- i \bar\chi_\beta \gamma^\beta  
\gamma^\alpha \partial_\alpha X_\nu \psi^\nu
\biggr],
\ea
where we have redefined
\[
\psi \to 2 \psi, \ 
\eta^{-1}\chi \to \bar\chi_\alpha\gamma^\alpha, \
\eta \to l_p \eta,
\]
where first Greek letters  run from $0$ to $D_\sigma - 1$, $D_\sigma$
again is the maximal number of the orthogonal directions toward 
which the uncertainty-caused propagation of a point particle takes place.
Averaging the seed action over all the deviations
$\sigma^i$ in the spirit of Eq. (\ref{eqMB}),
we eventually obtain the following supermetabrane action
\ba                                                \lb{eqsuMB}
&&S_{MB} = - \frac{1}{V_D}
\int [d \sigma] \sqrt{-\gamma}
\biggl(
\gamma^{\alpha\beta}
\partial_\alpha X^\nu \partial_\beta X_\nu    \nn\\
&& \qquad \qquad
+ 2 i \bar\psi^\nu \gamma^\alpha \partial_\alpha \psi_\nu
- i \bar\chi_\beta \gamma^\beta  
\gamma^\alpha \partial_\alpha X_\nu \psi^\nu
\biggr),
\ea
see the paragraph after Eq. (\ref{eqMB}) for corresponding notations.

It is straightforward to check that 
if the conditions near Eq. (\ref{eqGss})
are hold (assuming the analogous ones for the fermionic part) then
Eq. (\ref{eqsuMB}) yields 
the superstring 
\ba                                                
&&S_{S} = - \frac{1}{8\pi \alpha'}
\int d \sigma^0 d \sigma^1 \sqrt{-\gamma}
\biggl(
\gamma^{ab}
\partial_a X^\nu \partial_b X_\nu    
+\nn\\
&& \qquad \qquad
 2 i \bar\psi^\nu \gamma^a \partial_a \psi_\nu
- i \bar\chi_b \gamma^b  
\gamma^a \partial_a X_\nu \psi^\nu
\biggr),
\ea
and it is clear that it is possible to repeat 
all the way of Sec. \ref{s-ps-sc},
and eventually
obtain the similar picture with the only exception:
both the strings and, which is very important, 3-brane Universe
do contain a fermionic sector now.

Some final comments concerning the whole Sec. \ref{s-ps}
are in order.
One may ask the question what are the possible applications
of this viewpoint (in addition to the physically clear but abstract
construction justifying the use of extended objects in field theory).
It definitely hard to completely answer this question at this stage
but some significance can  be already outlined.

At first, this formalizm can be regarded as the unique procedure 
which is inverse to
worldvolume reduction - we start with a point-particle and
obtain branes of higher dimension including strings.
Therefore,
one may
regard the proposed approach as the realization of the
worldvolume holography principle.
In any case, it equips us with some kind of hierarchy (correspondence)
between branes of
different worldvolume dimensions.
In its turn, this correspondence can be used for obtaining of
valuable information about embeddings.
For instance, it is well-known that strings are renormalizable
at the quantum level whereas other branes are not,
therefore, one may do the following: study the string 
(classically or non-classically)
then uplift its worldvolume dimension by means of the proposed approach
so that the string will not be a string anymore and becomes
a brane but then a solid piece
of information about the new-born brane can be obtained from string theory.

At second, as was established above the simplest bosonic and fermionic 
particles correspond respectively to the simplest bosonic and fermionic
strings.
It is an exciting question what kind of strings corresponds to 
more complicated models of particles such as those with worldline
curvature, torsion (with higher derivatives, in general), 
with charges of several types, curved-space corrections, etc.

At third (but of course not actually last), 
the question which is inverse to the previous one -
what kind of particles  corresponds to 
the known five (I, IIA, IIB, heterotic)
strings - is also of large interest in its own.
Besides,
it may happen (by analogy with the quantum-mechanical Matrix 
theory \cite{bfss}) 
that different strings originate from the only certain
particle - this is, in fact, an alternative idea to that depicted 
in Fig. \ref{f-pp} where strings are the descendants of (meta)brane's 
``decay''.

\section{Conclusion}\lb{s-c}

Let us overview now the whole paper.
As was mentioned above it can be divided into the 
two parts which are independent of each other for a first look.
The first part (field-to-particle transition and its
applications, Secs. \ref{s-ea} - \ref{s-sg}) itself
was discussed in Secs. \ref{s-i} and \ref{s-d} so
here we will take a look on the part one 
only in relation to the part two.

\begin{figure}[htbp]
  \begin{center}
    \leavevmode
    \epsfxsize=0.9\columnwidth
    \epsfbox{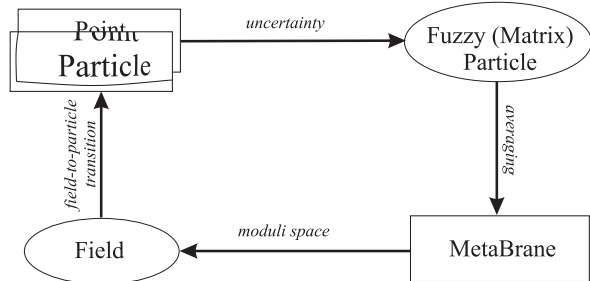}\vskip1mm
\caption{
Everything starts from the bare point particle;
inevitable uncertainty under macroscopical measurement 
breaks the
determinism and the particle becomes fuzzy (6.7), (6.28) 
from the viewpoint of a  measurer; 
after averaging over the deviation states
(in fact, this is what the procedure of a measurement does) 
we obtain the metabrane (6.8), (6.29)
which has a varying number of worldsheet dimensions;
the metabrane decays (see Fig. 1)
into the superposition of $\ast$-branes including strings:
the 
3-brane yields our Universe, strings produce fields
including gravity; 
eventually, physical field solutions after
field-to-particle transition 
result 
in the nonminimal point particle(s), see Secs. II-IV.
It is clear that the initial particle and the
resulting one (after a cycle)
are not the same - this is graphically emphasized 
by lens distortion, so the latter (whose action contains the 
higher-derivative terms) perhaps begins the new cycle.
The square boxes unlike the oval ones mean the  
deterministic embeddings.
}
    \label{f-ht}
  \end{center}
\end{figure}

In the discussion of the 
first part it was conjectured that the point particle
is not only the end product of field theory but also
the justification of the string-brane approach.
Then in the second part this conjecture was deepened: it turns out
that the real point particle from the viewpoint of a macroscopic
observation can
look like the extended object and hence can be effectively 
described in terms of branes and strings.
Then the 3-brane can be regarded as our Universe
whereas strings describe gauge symmetry of quantum field theory.
One may pictorially imagine the picture which unifies these two parts and
describes the role of each of them - Fig. \ref{f-ht}.

It is doubtless that a lot of work has to be done in future.
In connection with the part one:
the interesting question is 
what is the nonminimal noninteger-spin particle.
It seems that there exist two ways of how to treat it: 
(a) use the concept 
``noninteger-spin particles without spinors'' \cite{txt-nm,pl-sp}, or
(b) construct the models of supersymmetrical nonminimal particles.
The second task is to expand the field-to-particle transition
formalism on the cases of higher spacetime dimensions and
more complicated fields.
It seems that the whole picture holds but details may differ.

In connection with the part two: we have a lot of directions for further 
studying and a number of unclear points,
see the discussion at the bottom of Secs. \ref{s-ps-sc} and \ref{s-ps-fe}.
Besides, there exists a number of additional questions.
The first one is what is the number of dimensions of
the spacetime where the fundamental particle lives.
It is interesting that besides of the standard numbers $10$
and $11$ it was proposed (for the case
of a minimal fermionic particle) the number $14$ as a result of
the purely spacetime
realization of the symmetry group of the Standard Model
$SO(1,3) \times U (1) \times SU (2) \times SU (3)$ times an additional
extracharge $U (1)$ \cite{mar}.
The second question is about the dualities and metabrane's internal
content.
Once we try to include the strings into the one metabrane what are the
place and role of the string-string dualities?
Can they be regarded as the relations between the metabrane's
components?
Also, what is the explicit internal structure of the metabrane?

\section*{Acknowledgments}

I am grateful to 
Edward Teo (DAMTP, Univ. of Cambridge and Natl. Univ. of Singapore)
and 
Mikhail O. Katanaev (Steklov Math. Inst.) 
for helpful discussions concerning dilaton gravity and branes
as well as to
Belal E. Baaquie (Natl. Univ. of Singapore) for bringing
some papers into my attention and for constant encouragement
in superstring theory \cite{kb-l}. 
At last but not least, I acknowledge the correspondence by Mikhail
S. Plyushchay (Univ. de Santiago de Chile) enlightening some
subtle points of the higher-derivative particles' mechanics.

\def\AnP{Ann. Phys.}
\def\APP{Acta Phys. Polon.}
\def\ATMP{Adv. Theor. Math. Phys.}
\def\CJP{Czech. J. Phys.}
\def\CMPh{Commun. Math. Phys.}
\def\CQG {Class. Quantum Grav.}
\def\EPL{Europhys. Lett.}
\def\FP{Fortschr. Phys.}
\def\GRG {Gen. Relativ. Gravit.}
\def\IJMP  {Int. J. Mod. Phys.}
\def\JMP{J. Math. Phys.}
\def\JPh{J. Phys.}
\def\LMPh {Lett. Math. Phys.}
\def\MPL  {Mod. Phys. Lett.}
\def\NCim {Nuovo Cimento}
\def\NPh  {Nucl. Phys.}
\def\PhE  {Phys.Essays}
\def\PhL  {Phys. Lett.}
\def\PhR  {Phys. Rev.}
\def\PhRL {Phys. Rev. Lett.}
\def\PhRp {Phys. Rep.}
\def\TMF {Teor. Mat. Fiz.}
\def\TMP {Theor. Math. Phys.}
\def\prp {report}
\def\Prp {Report}

\def\jn#1#2#3#4#5{{#1}{#2} {\bf #3}, {#4} {(#5)}} 

\def\boo#1#2#3#4#5{{\it #1} ({#2}, {#3}, {#4}){#5}}



\begin{references}
\bibitem{txt-dil}
More precisely, in general case $T^\alpha$'s coincide 
with the gamma matrices up to the scalar-field coefficient which can
be associated with the square root of dilaton.
If so then the physical origin of the dilaton field is deeper than it
is usually regarded nowadays.
\end{references}
\end{document}